\documentclass[conference]{IEEEtran}
\IEEEoverridecommandlockouts
\usepackage{cite}
\usepackage{amsmath,amssymb,amsfonts}
\usepackage{algorithmic}
\usepackage{graphicx}
\usepackage{textcomp}
\usepackage{xcolor}
\usepackage{hyperref}

\usepackage[normalem]{ulem} 
\usepackage{xcolor}

\usepackage{ifthen}
\usepackage{amssymb}
\newboolean{showcomments}
\setboolean{showcomments}{true} 
\ifthenelse{\boolean{showcomments}}
  {\newcommand{\nb}[2]{
    \fcolorbox{gray}{yellow}{\bfseries\sffamily\scriptsize#1}
    {\sf\small$\blacktriangleright$\textit{#2}$\blacktriangleleft$}
   }
   
  }
  {\newcommand{\nb}[2]{}
   
  }

\usepackage{booktabs}
\usepackage{array}
\usepackage{colortbl}
\usepackage{multirow}
\usepackage{longtable}

\newcolumntype{v}[1]{>{\raggedright \hspace {0pt}}p{#1}}
\newcolumntype{G}[1]{>{\columncolor{gray90}}#1}

\definecolor{Gray}{gray}{0.8}
\definecolor{gray25}{gray}{0.25}
\definecolor{gray50}{gray}{0.50}
\definecolor{gray75}{gray}{0.75}
\definecolor{gray90}{gray}{0.9}

\newcommand{\interviewquote}[2]{\begin{quote}
\footnotesize{\emph{``#1'' }} --- \footnotesize{#2}
\end{quote}}

\def\BibTeX{{\rm B\kern-.05em{\sc i\kern-.025em b}\kern-.08em
    T\kern-.1667em\lower.7ex\hbox{E}\kern-.125emX}}
\begin{document}

\title{
%
Managing Human Factors in Automated Vehicle Development: Towards Challenges and Practices


}

\author{\IEEEauthorblockN{Amna Pir Muhammad, Eric Knauss}
\IEEEauthorblockA{\textit{Dept. of Computer Science and Eng.,}\\
\textit{Chalmers $\mid$ University of Gothenburg,} \\
Gothenburg, Sweden\\
Orcid: 0000-0001-8328-4149, {0000-0002-6631-872X}}

\and
\IEEEauthorblockN{Jonas B{\"a}rgman}
\IEEEauthorblockA{\textit{Dept. of Mechanics and Maritime Sciences,}\\
\textit{Chalmers University of Technology}\\
Gothenburg, Sweden\\
Orcid: {0000-0002-3578-2546}
}%
\and

\IEEEauthorblockN{Alessia Knauss}
\IEEEauthorblockA{\textit{Zenseact AB} \\
Gothenburg, Sweden\\
Orcid: {0000-0003-4857-7784}}

}

\maketitle

\begin{abstract}


 Due to the technical complexity and social impact, automated vehicle (AV) development challenges the current state of automotive engineering practice.
Research shows that it is important to consider human factors (HF) knowledge when developing AVs to make them safe and accepted.
This study explores the current practices and challenges of the automotive industries for incorporating HF requirements during agile AV development.
We interviewed ten industry professionals from several Swedish automotive companies, including HF experts and AV engineers.
Based on our qualitative analysis of the semi-structured interviews, a number of current approaches for communicating and incorporating HF knowledge into agile AV development and associated challenges are discussed. Our findings may help to focus future research on issues that are critical to effectively incorporate HF knowledge into agile AV development.

\end{abstract}

\begin{IEEEkeywords}
Requirements Engineering, Automated Vehicles, Human Factors, Agile Development
\end{IEEEkeywords}

\section{Introduction}

The development of automated vehicles (AVs) has the potential to revolutionize transportation and greatly improve safety, efficiency, and convenience. However, the successful deployment of AVs depends on its ability to effectively interact with human users and adapt to the complex and dynamic environments in which they operate \cite{seppelt2016potential}.
Many studies have identified human-related challenges in automation, such as over-trust and over-reliance \cite{Toshiyuki} of humans in AVs, misuse of the automation \cite{lee2009human}, interaction of drivers and AVs, transfer of control, and communication with other road users \cite{kyriakidis2019human}. 


Researchers argue that it is crucial to incorporate human factors (HF) knowledge into AV development to ensure their safety, usability, and acceptability \cite{John,Jordan}. Hence, HF knowledge must play a vital role throughout the development lifecycle of AV technology. 
Human factors refer to the physical, cognitive, social, and emotional components of individuals that can affect their performance and interactions with systems (such as AVs) \cite{hfes}.

Researchers emphasize the inclusion of human knowledge in the early stages of the design process in the requirements engineering phase \cite{haakansson2020including}. This allows for a more holistic and user-centered design process, where the needs and preferences of the human users are understood and considered from the beginning. 
%
Previously, in plan-driven development (e.g., based on V-model) HF research would have taken place in pre-development activities and would have contributed to up-front requirements specifications \cite{royce1987managing}. 

However, in recent years, there has been a shift towards agile development methods. Agile development focuses on rapid prototyping, continuous delivery, and frequent feedback and collaboration between self-organizing and cross-functional teams \cite{highsmith2002agile}. 
 Using agile development in the AV industry presents both opportunities and challenges to incorporate HF knowledge. Agile development allows for user feedback to be incorporated more quickly and flexibly and allows iterations on the design \cite{Meyer2014}. In contrast to these advantages, it is difficult to systematically derive the requirements of systems (such as AVs) in agile development \cite{Kasauli2020b}. For this reason, it is challenging to include HF knowledge in agile AV design \cite{zykov2020agile}.

{To bridge this gap and address the challenges, it is crucial to gain insights into how practitioners in the field actually include HF knowledge in their daily work and identify the specific obstacles they encounter. Previous studies have highlighted the need for addressing this issue \cite{muhammad2021methods}. By examining the practical experiences of professionals in the AV industry, we aim to shed light on the current practices and uncover the key challenges faced in integrating HF knowledge into agile AV design.}

 


To set the stage for more extensive investigations, in this exploratory  
study we represent a first step in the direction of exploring current practices and challenges for incorporating HF knowledge into the agile development of AVs. Using semi-structured interviews with ten industry experts (e.g., AV engineers, HF experts), this qualitative study aims to address the following research questions (RQs): 

\begin{description}
\item[RQ1:] What are the current practices to include HF knowledge in agile AV development?
\item[RQ2:] What practical challenges exist to deliver HF knowledge to developers in agile AV development?

\end{description}


Our findings may help to plan and conduct future studies. If the research community pays more attention to the needs and expertise of HF practitioners, we believe the industry will embrace new discoveries and work with researchers to better address HF in agile AV development.

{This paper builds upon interview data from a previous study, which examined the properties of agile development and HF and identified implications for HF, agile development, and requirements engineering. While the previous study aimed to explore and characterize these properties and implications, our current study focuses specifically on concrete practices and challenges.}
Key findings from the other paper can be found \href{https://doi.org/10.5281/zenodo.8094272}{here}.

The rest of the paper is divided into five sections. Section II reviews the background and Section III discusses the methodology. The findings are presented in Section IV, while Section V presents the discussion and threats to the validity of this study. Finally, Section VI concludes the paper.


\section{Related work}

{To our knowledge, there is only little related work about incorporating HF knowledge into agile AV development.}
{Thus}, we consider literature that deals with the integration of HF aspects into agile software development, as well as studies integrating HF aspects into AVs. Note that we do not focus on HF in terms of their impact on the employees of agile way of working as, for example, studied by \cite{lenberg2015human, chagas2015impact}.

\paragraph{Human Factors in Agile Software Development}
While studies provide guidelines for including HFs in the product development \cite{alan, mejdal2001human}, it does not specifically address the integration of HF requirements in agile development.
When it comes to incorporating HF into requirements engineering in agile development, some studies have focused on including HF in agile development \cite{bourimi2010affine, maguire2013using, pereira2018design, schon2016enterprise}.
The mentioned studies have limitations in scope and may not fully capture the complexity of HF aspects such as user research, usability testing, and feedback collection. They also lack empirical evidence to support their effectiveness in real-world agile projects and do not provide detailed guidelines for implementation. 
Saghafian et al. \cite{saghafian2021application} analyze the current level of integration of HF in immersive visual technologies (IVTs) development and highlight areas that need improvement. 
 This study also addressed the agile perspective, but then it is not for AVs and requirements perspectives. 


\paragraph{Human Factors in AV Development}
In recent years HF experts have made significant contributions to the understanding of how important HF are for AVs. Topics of relevance are interactions between AVs and the human driver \cite{kyriakidis2019human}, human engagement and disengagement\cite{Morando2021}, human-machine interfaces\cite{Kirsten}, situational awareness \cite{Chen}, etc. However, consideration of the knowledge of HF capabilities and limitations in these studies is limited to scenarios or specific design solutions for one of the aspects which traditionally would then be integrated into the requirements specification/design. However, it remains unclear how this kind of HF knowledge which is build up from research serves as input for agile AV development. 


\section{Method}

We used a qualitative study 
{design inspired by Maxwell} \cite{Maxwell} based on semi-structured interviews to explore
the current practices and challenges of incorporating HF in agile AV development. We chose this method due to the fact that this is a relatively unexplored research field and qualitative research would allow to explore the different aspects derived from individuals' insights based on their experience and understanding \cite{MADELEINE}.
 

\subsection{Data Collection}

\begin{table}
\caption{Interviewees’ roles and work experience (Experience level: Low= 0–5 years, Medium=5–10 years, High= More than 10 years)}
\label{tab:Participant}
\begin{center}
\begin{tabular}{lv{0.5\linewidth}v{0.25\linewidth}}
\toprule
\textbf{ID} & \textbf{Role} & \textbf{Experience Level  }
 \tabularnewline
 \midrule
 P1 & HF Expert (Specialist) 
 & High
  \tabularnewline
 P2 & HF Expert (Strategy, Specialist \& Research)
 & High
  \tabularnewline
  P3 & AV Engineer (Strategy \& Architecture) 
 &  High
  \tabularnewline
  P4 &  AV Engineer (Requirements \& Research)  
  & Medium
  \tabularnewline
  P5 & HF Expert (Management \& Research) 
  & High
  \tabularnewline
  P6 & HF Expert (Specialist) 
  &  High
  \tabularnewline
  P7 & HF Expert (Specialist \& Design) 
  & High
  \tabularnewline
  P8 & AV Engineer (Safety \& Research )
  & Low
    \tabularnewline
  P9 & AV Engineer (Strategy \& Specialist) 
  & High
    \tabularnewline
  P10 & HF/AV Engineer 
  & High
  \tabularnewline

\bottomrule
\end{tabular}
\end{center}
\end{table}

Our aim was to have the freedom to explore any new developing areas while maintaining a framework that enables the replication of our findings \cite{de2001research}. To achieve this, we conducted semi-structured interviews. Semi-structured interview approaches facilitate capturing intrinsic characteristics and allow the interviewer to probe further questions \cite{Creswell}.


For this study, we interviewed ten professionals (P1-P10) with diverse roles from various Swedish automotive companies such as Volvo (cars and trucks), Zenuity, Veoneer, and Autoliv. 
All interviewees have several years of experience working with AV development.
Table \ref{tab:Participant} presents the role of each participant along with their experience level.

Interviews lasted from 60 to 80 minutes. In most interviews, three authors were present during the interviews, while at least the first and second author were present in every interview. The first author was leading the interview, the second author taking notes. 
%
%
%
%
Before conducting the interviews, we created an interview guide to ensure that the same topics were covered in each interview. The guide included nine open-ended questions and detailed follow-up questions, which were covered in each interview. 
The interview questions used can be found \href{https://doi.org/10.5281/zenodo.8094272}{here}.

\subsection{Data Analysis}
In order to address the research questions (RQs), we thoroughly analyzed 
the responses to questions 3, 4, and 5 from the interview guide. However, to enhance the depth and scope of our data and findings, we also thoroughly examined and analyzed the entire data in relation to our research questions 
on the challenges and practices of incorporating HF in agile AV development.

While agile development was not initially a primary topic of the study when preparing for the interviews, our interviewees consistently provided answers that reflected the agile approach to working. As a result, we expanded our research questions to include the agile perspective.

For the qualitative analysis we used the thematic analysis approach \cite{Victoriac} to determine themes and analyze the content. Generally, this approach consists of six steps.
 %
 Following the steps, first we reviewed all interview notes thoroughly and created memos on topics relevant to our research questions. Then the text was assigned labels called codes, using both Microsoft Word and Nvivo. The coding scheme was refined iteratively to uncover important ideas and perspectives. 
The codes were then analyzed and grouped to find common patterns to define the themes. 
The themes that emerged through the coding were then reanalyzed and checked to identify any ambiguous, contradicting, or missing ideas. 
 Finally, we expanded on the descriptions of the themes and provided example quotes.
The themes and findings are presented in the next section.

\section{Findings}

This study aims to investigate the current practices on how HF knowledge is currently incorporated in agile AV development and what challenges are there.

\subsection{Current Practices}

In this section, we report current practices which practitioners use to include HF knowledge in agile AV development (RQ1).

\paragraph{\textbf{Derivation of HF Requirements}}
Our interviewees indicated that there is just no one specific method to derive HF requirements. 
Rather the process varies both between companies and depending on the scope of the HF being considered.

While only a few of our interviewees were able to provide specific details on how they derive and collect HF requirements, some examples included, using statistical data (e.g., to understand the concrete context or specific driver behavior), getting specifications from customers (original equipment manufacturers (OEMs)), using prototypes (to understand human interaction),  and conducting research.

\interviewquote {A lot of requirements are derived from research or common HF knowledge as a requirement bible for driver interface. That is the big list of all requirements. But you cannot just say that everything in there just applies because it depends on the context choice. But we have that as a basis.} {P7}


In addition, in some cases, user studies are also carried out to find the HF requirements — through co-driving with test drivers or meeting them in clinics. 

\interviewquote{In addition, user studies to meet people. For each new function, for example, that we develop. This might be through co-driving with test drivers or meeting them in clinics. Finding the requirements and also validating them.} {P7}




Regarding human-machine interaction (HMI), some HMI/HF experts get involved in writing requirements.  

{\interviewquote{-- is in charge of HMI and interaction. So, he is writing early requirements, when initiated by our product planning. He is also following up, testing, and supporting the refinement.}{P6}}

\paragraph{\textbf{Communication and Documentation of HF Requirements}}
According to our interviewees, requirements are traditionally communicated to others through documents.
In agile development, however, teams rely more on informal, regular meetings and interactions to align on requirements. In this context, it is not clear how to effectively transfer this knowledge between multiple domain experts.

Different companies use different tools and methods depending on the context to facilitate knowledge transfer of requirements in agile development. For example, it varies depending on the situation or type of information and to whom it is provided. Our interviewees mentioned using various tools to communicate and document requirements, including, for example, JIRA, some requirements tools, or PowerPoint.







Some interviewees stated that the best way to communicate HF knowledge to AV developers is to make HF part of the teams. They indicated that otherwise, it will be challenging to incorporate HF knowledge effectively if HF experts are not part of the team. However, the lack of HF specialists makes it difficult to have them integrated in each team.

\interviewquote {How would HF experts bring up their knowledge? If the network exists, just a quick call over Skype. If not: You are put into the same project together. If the project has been implemented and you keep getting reports on features, I do not know if that would first go to HF or engineers, but either way, one would contact the other.} {P9}


\paragraph{\textbf{Consideration of HF}}

There were different views about the current practices for consideration of HF knowledge in developer's daily work. Most interviewees expressed that HF expertise is included mainly in the HMI development. Others believed they do not have a distinct division or allocation of responsibility for HF inside their organizations.

When it comes to HMI or user experiences, in some companies, a HF expert is there to incorporate HF knowledge and ensure that HF knowledge is covered. However, in the absence of HF experts, HMI designer or user experience (UX) engineers perform the tasks and is in charge of incorporating HF into their designs. 


\interviewquote{For me that is a lot of walking. You need to visit a lot of parts of the company premises to meet people.}{P7}


Lastly, most of the interviewees mentioned that there is not enough consideration of HF in development, which is particularly affected by the short sprints of agile development and also because of the lack of HF experts.


\paragraph{\textbf{Interaction and Communication with AV Developers}}

Most of the interviewees expressed that there is almost no or infrequent direct communication between HF experts and AV developers. 

\interviewquote{ Very rarely, unfortunately... I can talk to their group leader, but indirectly.} {P5}

However, considering HF knowledge in HMI, we learned through this study that a few communications occur where teams try to have an HMI specialist in each agile team. These HMI experts then work with the product owner (PO) or product managers (PM) to create user stories and develop features. It is, however, still tricky to execute experiments from HF perspectives.

\interviewquote{But we write the backlog items in ADAS. That works well. We have a few of us that hook into ADAS stream, but many other HF are only in the HF stream.}{P6}



 \paragraph{\textbf{Experiments and Tests with HF Knowledge}}
 In the experience of our participants, HF experiments can offer valuable insights into designing products that are more efficient, user-friendly, safe, and optimized for human performance. However, only a small number of participants mentioned actually conducting such experiments, and even then, only for new functionalities. Typically, these studies involve co-driving with test drivers, user studies, or meeting with people in clinics.



{\interviewquote{ For each new function, for example, that we develop. This might be through co-driving with test drivers or meeting them in clinics. Finding the requirements and also validating them.}{P7}}



\subsection{Challenges}

This section presents the challenging aspects of integrating HF knowledge to AV developers in agile development that we identified in our interview study (RQ2).

\paragraph{\textbf{Communication}}
Most of our interviewees pointed out that communicating HF knowledge in the current ways of agile working is difficult for several reasons. For example, currently, there is no defined methodology for how or where HF should be included in the team and development lifecycle. Our interviewees mentioned that it would be easier to convey HF knowledge if HF specialists were part of the team. 
It is also hard to integrate HF in the workflow of developers in agile development because of short sprint cycles. Agile development is a fast-paced process that demands fast decisions, thus it can be challenging to have enough time to integrate and communicate HF knowledge.

\interviewquote{ Yes, because we are very big, and we may not have the time or even the amount of information is too big to catch up in the right timeframe.}{P3}
 

 \interviewquote{The way you communicate your requirements is within the teams. You need to be there. If you are not in the teams, it will be a challenge.} {P7}

\paragraph{\textbf{HF Experts not Part of Teams}}

 According to our interviewees, it is difficult to include HF in agile AV development because agile teams are autonomous and responsible for discovering knowledge by themselves and, therefore, must have knowledge of HF. However, many of these teams lack the necessary knowledge and expertise in HF. 
 
 \interviewquote{They(teams) are then responsible for the topic. T-shaped teams. But we are lacking HF people.}{P2}
 
This is because HF is a specialized subject that demands specialized knowledge and skills, which engineers generally do not have.

 \interviewquote{We also try to take part during the development. This agile way of working, we figure that out ourselves. We have tried different things. We had one HMI expert in each team, but that did not scale. We do not have enough experts to have one in each team for 100\%.} {P7}

In the absence of HF experts in the development team, there may be a lack of understanding of specific HF issues that are critical to the development of AVs. This can result in missing important design factors that can affect user experience, usability, and safety.
%
One reason for the lack of HF in the development team is that organizations currently do not have enough HF experts.

\interviewquote {The problem is, if you are not on the train, you are not able to promote yourself. If you are a shared resource team, you have less visibility. So it will be better to be on the train.} {P6}




\paragraph{\textbf{Knowledge Management}}

Another challenge that has been identified is the lack of a proper way to manage the knowledge, for example, how to store this knowledge for future use, and how to use the knowledge accumulated from previous projects. Our interviewees mentioned that without a framework or model, it is difficult to collect and use lessons learned from previous projects and apply that knowledge to current development activities. As a result, inefficiencies can occur, opportunities for improvement can be missed, and progress can slow down.

\interviewquote {We could have high-level requirements: Anything that the vehicle does should not surprise other road users and other similar requirements. We are not yet very good at managing such requirements. }{P7}

Additionally, without a common vocabulary or framework for discussing and sharing this knowledge, it is difficult to communicate effectively and ensure that HF aspects are properly incorporated into the development process. 
Our interviewees stated that the key to addressing these challenges is to establish unique frameworks and models for managing the knowledge of human elements in AV creation. These frameworks should provide a mechanism for defining and capturing needs in terms of HF, as well as incorporating these needs into the general development process.


\interviewquote {At the moment, I am not even sure what we mean by these requirements. A framework is missing. HF models are not completely in place, so it is very hard to say if we can convey the requirements in a complete way. } {P8}

\paragraph{\textbf{HF Experiments and Tests}} 
Several participants in the study highlighted the significant importance of HF in field experiments and testing of AVs. However, they also highlighted that conducting HF experiments is challenging for several reasons. Most notably, such experiments can be both expensive and time-consuming, which can be particularly challenging in agile development environments. 
This is because agile methodologies prioritize rapid iteration and feedback, which can make it difficult to conduct a thorough examination of HF issues to ensure vehicle safety and efficacy.
 Striking a balance between the rapid iteration and feedback of agile development and the thorough examination of HF issues can be challenging.

Moreover, the study of HF in AVs is still in its early stages of development. Unlike other fields, such as software testing, where tests are formal, established, and mature, HF tests are not yet formalized enough for automated testing.

\interviewquote{Established, mature tests are very much formalized, but we are lacking that formality.}{P2}



Additionally, our interviewees mentioned that tests and simulations can present challenges when attempting to make a complete safety argument. Therefore, it may be more effective to argue based on requirements rather than attempting to create a comprehensive safety argument based on testing alone.

\interviewquote{Tests and simulation are troublesome when attempting to make an argument for completeness. If one aims for a complete safety argument, it might be better to argue based on requirements.} {P8}





\paragraph{\textbf{Mindset}}

Some participants in the study emphasized that applying HF knowledge to agile development requires more than just adopting new practices. This requires a cultural change within the organization and a shift in developers' mindsets.
However, changing culture and mindset is a complex and long-term process, making it challenging to implement.

\interviewquote{If you do not have the company philosophy, it is hard to make the connection. It is much more than describing the path toward a specific requirement in a database.} {P1}

 To promote HF knowledge in agile AV development, interviewees suggested that organizations and developers should focus on changing their culture and mindsets rather than solely scaling HF experts or developing better frameworks. By incorporating HF into their daily work, they can create a culture of awareness and understanding of the importance of HF in AV development.


\interviewquote{The reason that HF is overlooked since it is a traditional engineering company. They like to bring in a UX engineer rather than work on the mindset.} {P6}

\section{Discussion and Threats to Validity}

The study findings suggest that determining HF requirements is not a one-size-fits-all process and varies depending on the organization and team. Practitioners should be aware of different approaches to identifying needs and select the most suitable one based on their specific circumstances. Currently, HF knowledge is primarily concentrated on HMI development, but as AV behavior encompasses features beyond HMI, there is a need to expand the integration of HF knowledge.

Incorporating HF expertise poses challenges, as also indicated by \cite{madni2011integrating, saghafian2021application}. HF experts should be part of the development team to effectively share their knowledge and contribute to design decisions. This suggests that organizations should consider restructuring their teams to include HF specialists, improving knowledge transfer and timely communication of HF requirements. Furthermore, the lack of tools and procedures to support HF in agile AV development presents another obstacle. Therefore, standardization and streamlined setups for executing HF experiments and tests are necessary.


We also find that there is a need to work on the developers' mindset in order to effectively bring HF knowledge to development.
Engineers should understand the importance of HF knowledge and have enough awareness to know when to seek out HF expertise and how to apply HF knowledge in development. 
Our findings also {confirm} that HF should be given greater consideration in requirements {(as for example discussed by Håkansson et al. }\cite{haakansson2020including}) and testing.



\subsection{Threats to Validity}
We consider the four perspectives of threats to validity   covered by Easterbrook et al. \cite{easterbrook2008selecting}. 

\paragraph{Construct Validity} 
One of the most important aspects of validation in qualitative research is construct validity. 
The risk of construct validity was addressed by including authors who have sound knowledge and experience working with automotive industry and HF.
The last three authors of the study have extensive experience in requirements engineering and HF within the automotive industry from research and development perspective.
They assisted the first author by leveraging their expertise in developing an interview guide that accurately reflected the research objectives of the study. Additionally, the interview guide was refined through multiple iterations to ensure its clarity and relevance.

\paragraph{Internal Validity} 
To minimize this risk, we took great care to collect data about the topics and their contexts and provided detailed descriptions of our findings. We used data triangulation and interviewed multiple roles and people in each role to ensure internal validity. However, there may still be a selection bias because the interviews were selected through our industry contacts. To ensure continuity in the data collection process, we conducted all interviews using the same guide, but follow-up questions varied depending on the context.

\paragraph{External Validity} 
Our results are difficult to generalize due to the fact that all participants were working in Sweden and we focused on automotive companies. Thus, our results may not be applicable to other countries or other domain working with agile.
Despite the global presence of our interviewees' companies, cultural differences could still exist, impacting practitioners' reasoning about current practices and challenges for incorporating HF in agile AV development.
However, these findings are likely to provide valuable insights for HF experts and automotive companies interested in incorporating HF in agile AV development. It will also be useful for other researchers in the automotive domain.

\paragraph{Reliability} 

To minimize these risks, we took several measures. During the interviews, we had multiple researchers present to ensure the reliability of our data. We also share the material we used when conducting the interviews and analyses, which can be utilized by other researchers to replicate our methodology in various settings.
Additionally, the coding results were discussed among all authors to ensure consistency. Nonetheless, we recognize that some degree of subjectivity may still be present in our analysis.

\section{Conclusion and Future work}

The objective of this study is to explore current practices and challenges encountered by practitioners in incorporating HF knowledge into agile AV development. Our results imply that HF knowledge has not been well integrated by developers in 
{on the full AV system scope, but too often limited to the HMI design}. 
Challenges for incorporating HF in agile AV development include, for example, the lack of HF experts, knowledge management, and tools and procedures to properly engage HF experts. 
{Our results encourage developing a} robust setup with protocols to effectively incorporate HF knowledge and perform user studies so that developers can generate knowledge and data with good reliability.
%
We expect our results to be useful for engineers and managers aiming to improve the integration of HF knowledge in agile systems development.

\textit{Future Work:}
Our study shows clearly that more research is needed in this critical area and we hope that our initial results can contribute to this discourse.
{By expanding this study, we aim to collect best practices from AV development on how to integrate HF knowledge and requirements into large-scale agile workflows and to connect it to a organization-wide requirements strategy, e.g. based on \cite{muhammad2022defining}.}

{Our future work will focus on the integration of HF requirements within the agile workflow, for example, examining their incorporation in each sprint and daily work. We will investigate the roles and responsibilities surrounding HF requirements, including who is responsible for their implementation, and explore the methods used to store and manage these requirements}. 
{To gain a more comprehensive understanding, we plan to extend our qualitative survey to involve a broader range of companies and participants. This will enable us to capture a diverse set of perspectives and gather a more robust set of data on practices and challenges.}

{Furthermore, we intend to conduct an in-depth case study in an automotive company to develop a concrete requirement strategy that addresses the integration of HF knowledge and requirements across organizational boundaries. This is particularly important as many requirements originate from OEMs and are passed on to suppliers.}

{Through these research endeavors, we aim to contribute to the development of effective strategies and guidelines for integrating HF knowledge and requirements into large-scale agile workflows in the AV industry. Our ultimate goal is to enhance AVs' safety, usability, and overall quality by bridging the gap between HF considerations and agile development practices.}

\bibliographystyle{splncs04}
\bibliography{name}

\begin{thebibliography}{10}
\providecommand{\url}[1]{\texttt{#1}}
\providecommand{\urlprefix}{URL }
\providecommand{\doi}[1]{https://doi.org/#1}

\bibitem{bourimi2010affine}
Bourimi, M., Barth, T., Haake, J.M., Uebersch{\"a}r, B., Kesdogan, D.: Affine
  for enforcing earlier consideration of nfrs and human factors when building
  socio-technical systems following agile methodologies. In: Human-Centred
  Software Engineering: Third International Conference, HCSE 2010, Reykjavik,
  Iceland, October 14-15, 2010. Proceedings 3. pp. 182--189. Springer (2010)

\bibitem{chagas2015impact}
Chagas, A., Santos, M., Santana, C., Vasconcelos, A.: The impact of human
  factors on agile projects. In: 2015 Agile Conference. IEEE (2015)

\bibitem{Chen}
Chen, J.Y.C., Procci, K., Boyce, M., Wright, J., Garcia, A., Barnes, M.:
  Situation awareness-based agent transparency. ARL-TR-6905, Aberdeen Proving
  Ground, MD, U.S. Army Research Laboratory  (2014)

\bibitem{Victoriac}
Clarke, V., Braun, V., Hayfield, N.: Thematic analysis. Qualitative psychology:
  A practical guide to research methods  (2015)

\bibitem{Creswell}
Creswell, J.W., Creswell, J.D.: Research design: Qualitative, quantitative, and
  mixed methods approaches. Sage publications (2017)

\bibitem{de2001research}
De~Vaus, D.: Research design in social research. Research design in social
  research pp. 1--296 (2001)

\bibitem{easterbrook2008selecting}
Easterbrook, S., Singer, J., Storey, M.A., Damian, D.: Selecting empirical
  methods for software engineering research. Guide to advanced empirical
  software engineering pp. 285--311 (2008)

\bibitem{haakansson2020including}
H{\aa}kansson, E., Bjarnason, E.: Including human factors and ergonomics in
  requirements engineering for digital work environments. In: 2020 IEEE First
  International Workshop on Requirements Engineering for Well-Being, Aging, and
  Health (REWBAH). pp. 57--66. IEEE (2020)

\bibitem{highsmith2002agile}
Highsmith, J.A., Highsmith, J.: Agile software development ecosystems.
  Addison-Wesley Professional (2002)

\bibitem{hfes}
{Human Factors and Ergonomics Society}: Definitions of human factors and
  ergonomics (2021),
  \url{https://www.hfes.org/About-HFES/What-is-Human-Factors-and-Ergonomics},
  accessed on 17 Feb, 2021

\bibitem{Toshiyuki}
Inagaki, T., Itoh, M.: Human’s overtrust in and overreliance on advanced
  driver assistance systems: a theoretical framework. International journal of
  vehicular technology  \textbf{2013} (2013)

\bibitem{Kasauli2020b}
Kasauli, R., Knauss, E., Horkoff, J., Liebel, G., de~Oliveira~Neto, F.G.:
  Requirements engineering challenges and practices in large-scale agile system
  development. Journal of Systems and Software  \textbf{172} (2021)

\bibitem{kyriakidis2019human}
Kyriakidis, M., de~Winter, J.C., Stanton, N., Bellet, T., van Arem, B.,
  Brookhuis, K., Martens, M.H., Bengler, K., Andersson, J., Merat, N., et~al.:
  A human factors perspective on automated driving. Theoretical issues in
  ergonomics science  \textbf{20}(3),  223--249 (2019)

\bibitem{John}
Lee, J.D.: Humans and automation: Use, misuse, disuse, abuse. HUMAN FACTORS,
  Vol. 50, No. 3, p. 404–410 (2008)

\bibitem{lee2009human}
Lee, J.D., Seppelt, B.D.: Human factors in automation design. Springer handbook
  of automation pp. 417--436 (2009)

\bibitem{MADELEINE}
Leininger, M.: Evaluation criteria and critique of qualitative research
  studies. Critical issues in qualitative research methods 95  (1994)

\bibitem{lenberg2015human}
Lenberg, P., Feldt, R., Wallgren, L.G.: Human factors related challenges in
  software engineering--an industrial perspective. In: 2015 ieee/acm 8th
  international workshop on cooperative and human aspects of software
  engineering. pp. 43--49. IEEE (2015)

\bibitem{madni2011integrating}
Madni, A.M.: Integrating humans with and within complex systems. CrossTalk
  \textbf{5} (2011)

\bibitem{maguire2013using}
Maguire, M.: Using human factors standards to support user experience and agile
  design. In: 7th International Conference, UAHCI 2013. pp. 185--194. Springer
  (2013)

\bibitem{Maxwell}
Maxwell, J.A.: Qualitative research design: An interactive approach. Sage
  publications  (2012)

\bibitem{mejdal2001human}
Mejdal, S., McCauley, M.E., Beringer, D.B., et~al.: Human factors design
  guidelines for multifunction displays. Tech. rep., United States. Department
  of Transportation. Federal Aviation Administration~… (2001)

\bibitem{Meyer2014}
Meyer, B.: Agile! Bertrand MeyerThe Good, the Hype and the Ugl. Springer (2014)

\bibitem{Morando2021}
Morando, A., Gershon, P., Mehler, B., Reimer, B.: A model for naturalistic
  glance behavior around tesla autopilot disengagements. Accident Analysis \&
  Prevention  \textbf{161},  106348 (2021)

\bibitem{muhammad2021methods}
Muhammad, A.P.: Methods and guidelines for incorporating human factors
  requirements in automated vehicles development. In: REFSQ Workshops (2021)

\bibitem{muhammad2022defining}
Muhammad, A.P., Knauss, E., Batsaikhan, O., Haskouri, N.E., Lin, Y.C., Knauss,
  A.: Defining requirements strategies in agile: A design science research
  study. In: Product-Focused Software Process Improvement: 23rd International
  Conference, PROFES 2022. Springer (2022)

\bibitem{Jordan}
Navarro, J.: A state of science on highly automated driving. Theoretical Issues
  in Ergonomics Science  \textbf{20}(3),  366--396 (2019)

\bibitem{pereira2018design}
Pereira, J.C., de~FSM~Russo, R.: Design thinking integrated in agile software
  development: A systematic literature review. Procedia computer science
  \textbf{138},  775--782 (2018)

\bibitem{alan}
Poston, A.: Guidance on the application of human factors to consumer products.
  Tech. rep., Division of Human Factors, U.S. Consumer Product Safety
  Commission, Rockville, MD USA and Risk Assessment Division, Consumer Product
  Safety Directorate, Health Canada (2018),
  \url{https://www.cpsc.gov/s3fs-public/HF-Standard-Practice-Draft-12Feb2018.pdf},
  online (assessed: 2023-Mar-16)

\bibitem{Kirsten}
Revell, K., Langdon, P., Bradley, M., Politis, I., Brown, J., Stanton, N.: User
  centered ecological interface design (uceid):a novel method applied to the
  problem of safe and user-friendly interaction between drivers and autonomous
  vehicles. Intelligent Human Systems Integration,Advances in Intelligent
  Systems and Computing  (2018)

\bibitem{royce1987managing}
Royce, W.W.: Managing the development of large software systems: concepts and
  techniques. In: Proceedings of the 9th international conference on Software
  Engineering. pp. 328--338 (1987)

\bibitem{saghafian2021application}
Saghafian, M., Sitompul, T.A., Laumann, K., Sundnes, K., Lindell, R.:
  Application of human factors in the development process of immersive visual
  technologies: challenges and future improvements. Frontiers in psychology
  \textbf{12},  634352 (2021)

\bibitem{schon2016enterprise}
Sch{\"o}n, E.M., Winter, D., Uhlenbrok, J., Escalona~Cuaresma, M.J.,
  Thomaschewski, J.: Enterprise experience into the integration of
  human-centered design and kanban. In: ICSOFT-EA 2016: 11th International
  Joint Conference on Software Technologies (2016), p 133-140 (2016)

\bibitem{seppelt2016potential}
Seppelt, B.D., Victor, T.W.: Potential solutions to human factors challenges in
  road vehicle automation. In: Road vehicle automation 3, pp. 131--148.
  Springer (2016)

\bibitem{zykov2020agile}
Zykov, S.V., Singh, A.: Agile Enterprise Engineering: Smart Application of
  Human Factors, vol.~175. Springer (2020)

\end{thebibliography}

\end{document}